\documentclass{article}

\usepackage{amsmath}
\usepackage{amsfonts}
\usepackage{amssymb}
\usepackage{graphicx}
\usepackage{algorithmic}
\usepackage{algorithm2e}
\usepackage{cite}
\usepackage{xcolor}
\title{A Novel Supervisory Control Algorithm to Avoid Deadlock in a Manufacturing System Based on Petri Net in Presence of Resource Failure}
\author{
    Ahmad Bagheri\\
    Department of Electrical Engineering \\
    Amirkabir University of Technology \\
    Tehran, Iran \\
    \texttt{ahmadb007@aut.ac.ir} \\
    \and
    Mohammadhossein Aghaazizi \\
    Department of Electrical Engineering \\
    Amirkabir University of Technology \\
    Tehran, Iran \\
    \texttt{mohammadhosein79@aut.ac.ir} \\
    \and
    Ali Doustmohammadi \\
    Department of Electrical Engineering \\
    Amirkabir University of Technology \\
    Tehran, Iran \\
    \texttt{dad@aut.ac.ir}
}

\date{}

\begin{document}

\maketitle

\begin{abstract}
It is well established that resource failure, including robots and machines, in a manufacturing system can result in deadlocks. This issue not only hampers the system's performance but can also inflict significant damage on the manufacturing process. In this paper, we present a new algorithm developed through modeling of a manufacturing system using Petri net that ensures the liveness of the net in the event of such a failure. To detect possible failures, we first design a recovery subnet that is integrated into the resource. Next, we analyze the effects of failures on each state of the network to identify forbidden states. Finally, we propose an algorithm that optimally adds control places and establishes new constant vectors within the network, enabling effective management of remaining resources across different parts of the net. The proposed algorithm has been implemented in a system featuring three manufacturing lines, demonstrating its error-free operation while ensuring key properties such as boundedness, liveness, and performance continuity within the net.
\end{abstract}


\section*{Introduction}
In recent decades, various approaches have aimed at developing robust supervisory control systems. Lawley and Sulistyono \cite{b1} were pioneers in this field, designing a robust supervisory control policy that integrates a modified Banker's algorithm with neighborhood constraints for automated production systems. Their approach ensures that when a failure occurs in an unreliable resource, processes not dependent on that resource can continue operating, effectively preventing deadlock and blocking situations. Hao Yu et al. \cite{b2} further advanced the Banker's algorithm by targeting two key control objectives for shared resource systems with multiple unreliable components. Firstly, they ensured that parts of the network unaffected by the failure could continue functioning. Secondly, they developed mechanisms to facilitate the completion of remaining tasks after the repair of faulty resources. Feng et al. \cite{b3} introduced the concept of "Transition Cover-based robust controller" to create a robust controller for automated manufacturing systems under multiple unreliable sources. They subsequently designed an enhanced controller aimed at improving the features of their initial design, under the assumption that at most one unreliable source may fail at any given time. Recognizing the challenges associated with calculating all network siphons, Nan Do et al. \cite{b4} employed algebraic methods to present an iterative algorithm for identifying network siphons in systems with unreliable resources. Their approach included adding a control place to prevent siphons from emptying, thereby avoiding deadlock during resource failures. Across these methodologies, the overarching goal remains to maintain maximum network efficiency following a failure. Despite significant advancements in this area, few studies have concentrated on optimal robust controller design, as achieving this requires comprehensive calculations of all available states. However, as system size increases, the number of operational modes grows exponentially, making it increasingly difficult—if not impossible—to compute all possible states. In this context, Benyuwan Yang and Hesuan Hu \cite{b5} proposed an algebraic algorithm based on integer linear optimization that categorizes all system states into robust and non-robust states. A robust state is defined as one in which components not reliant on an unreliable source can continue functioning during a failure.

\section*{Definitions and preliminary concepts}
In the 1960s, Carl Adam Petri introduced the concept of Petri net to model and analyze complex systems.\cite{b6} Mathematically, a Petri net is represented as a 4-tuple in the form \( N = \langle P, T, F, W \rangle \), where \( P \) denotes a finite set of places, \( T \) represents a finite set of transitions, \( F \) is a set of arcs (\( F \subseteq (P \times T) \cup (T \times P) \)), and \( W \) is a weight function for the arcs. \cite{b10}

Liveness in a Petri net model indicates that at any stage, the system can transit to a new arrangement of markings by executing a sequence of transitions, with this process continuing indefinitely. In essence, ensuring the liveness property guarantees that in every state, at least one transition is enabled and can fire to move the network to the next state.

The boundedness attribute specifies whether there is a limit to the number of tokens that can fit in any places and in any given marking. In the absence of this limitation, tokens may accumulate in certain places during the network's evolution, potentially disrupting system operations. In practical terms, this implies that production components may accumulate in the system without any restrictions in buffers. However, given the finite capacity of each storage, prolonged accumulation can lead to overflow. Consequently, the boundedness characteristic ensures an upper limit on the number of tokens in each place, contributing to the proper and efficient functioning of the system.

The optimization of resources, increased flexibility, improved coordination, and reduced production time are some of the primary motivations for utilizing shared resources in production systems. Conversely, competition between processing lines for these limited resources, along with potential resource failures, can result in deadlock.\cite{b11}

A network reaches deadlock when one or more processes are indefinitely waiting for the release of resources currently held by other processes. A forbidden state is any state that breaches specific control requirements, leading to costly errors such as resource interference. Resource conflict arises when two or more processes simultaneously demand access to the same resource.

Supervisory control employs a supervisor controller to ensure the system operates within defined constraints and specifications. Deadlock control strategies generally fall into three categories: deadlock detection and correction, deadlock prevention, and deadlock avoidance.

The \( \text{S}^3\text{PR} \) network builds upon the \( \text{S}^2\text{PR} \) concept by adopting a system-level perspective where several simple sequential processes are interconnected. In such networks, each processing level can use at most one resource unit at any given state. The \( \text{S}^4\text{PR} \) network, an extension of the \( \text{S}^3\text{PR} \) model, addresses this limitation by enabling simultaneous and flexible resource utilization without altering the process model. This allows resources to be reused by other processes, embodying the property of conservativeness.

Overall, \( \text{S}^n\text{PR} \) networks are widely used to model shared resources. These networks consist of multiple sets of state machines and resources. Analyzing their evolution reveals that the \( \text{S}^2\text{PR} \) structure is the simplest form designed to allocate resources to a single process.\cite{b13}

In a Petri net, an S\textsuperscript{4}PR net is represented as: \cite{b7}
\begin{equation}
N = \sum_{i=1}^n N_i = (P, T, F, W)
\end{equation}
where:
\begin{equation}
N_i = (P_{A_i} \cup \{P_i^0\} \cup P_{R_i}, T_i, F_i, W_i), \quad i \in N_n.
\end{equation}

The set of operation places is defined as:
\begin{equation}
P_A = \bigcup_{i=1}^n P_{A_i},
\end{equation}
obtained as the union of all operation places across the subnets. Notably, the operation places within each subnet are exclusive, meaning no two subnets will share the same operator place:
\begin{equation}
\forall i, j \in N_n, \, i \neq j \implies P_{A_i} \neq \emptyset, \, P_{A_i} \cap P_{A_j} = \emptyset.
\end{equation}

Each subnet has a unique idle place, and the overall set of idle places in the S\textsuperscript{4}PR network is formed as:
\begin{equation}
P^0 = \bigcup_{i=1}^n \{P_i^0\}.
\end{equation}

In the modeling of automated production systems, each subnet is a fully connected state machine representing the production of a product. These subnets use resources to optimize the production process. The primary aim is to allocate resources efficiently to maximize network performance while utilizing available capacities. The set of resource places in this network is represented as in relation:
\begin{equation}
P_R = \bigcup_{i=1}^n \{P_{R_i} = \{r_1, r_2, \dots, r_m\} \mid m \in \mathbb{N}^+\}.
\end{equation}

The transitions of the network, represented as:
\begin{equation}
T = \bigcup_{i=1}^n T_i,
\end{equation}
are derived from the union of all subnet transitions. Similar to operation places, no two subnets will share a transition:
\begin{equation}
\forall i, j \in N_n, \, i \neq j \implies T_i \neq \emptyset, \, T_j \neq \emptyset, \, T_i \cap T_j = \emptyset.
\end{equation}

Each resource place \( r \) in the network is associated with a unique \( P \)-semiflow, denoted as \( I_r \), and is defined as follows:
\begin{equation}
\begin{cases} 
\forall r \in P_R \implies \{r\} = \|I_r\| \cap P_R  \\ 
\{p_0\} \cap \|I_r\| = \emptyset \\ 
P_A \cap \|I_r\| \neq \emptyset, \, I_r(r) = 1  
\end{cases}
\end{equation}

This relationship also specifies the number of operator places required for each resource, which can be calculated using the formula:
\begin{equation}
P_A = \bigcup_{r \in P_R} (\|I_r\| \setminus \{r\}).
\end{equation}

Each subnet:
\begin{equation}
N_i = (P_{A_i} \cup \{P_i^0\}, T_i, F_i, W_i), \, i \in N_n,
\end{equation}
excluding resource places, is a fully connected state machine, where each cycle includes a corresponding idle place.

Considering:
\begin{equation}
N = (P_A \cup P^0 \cup P_R, T, F, W),
\end{equation}
as an S\textsuperscript{4}PR network, the initial condition is valid if and only if:
\begin{equation}
\begin{cases}
\forall i \in N_n \implies M_0(P_i^0) > 0, \\
\forall p \in P_A \implies M_0(p) = 0, \\
\forall r \in P_R \implies M_0(r) \geq \max \{I_r(p) \mid p \in P_A\}.
\end{cases}
\end{equation}

It is evident that when a resource fails and leaves the network, the remaining resources experience reduced capacity, which directly impacts the network’s operation. Consequently, it is necessary to analyze the relationship between resource failures and network conditions. The remaining resources must then be redistributed across the network to sustain the maximum production rate.

\section*{Supervisory Control Using GMEC in S\textsuperscript{4}PR Systems}

Supervisory control is a higher-level system or agent tasked with overseeing and managing the operations of a subordinate system or process. A typical example can be observed in manufacturing facilities, where control rooms monitor the activities of various machines and processes. Operators in these control rooms have the ability to make adjustments, troubleshoot problems, and implement decisions to ensure that the system functions efficiently and securely. Essentially, effective monitoring and control of an application often require additional actions and modifications to the application itself.

Within a Petri net framework, these necessary adjustments are made by integrating one or more monitor places into the network as dictated by the specific problem at hand. For instance, in the accompanying figure, the addition of place \(p_c\) as a supervisory monitor restricts more than one token from entering \(p_2\), as it prevents transition \(t_1\) from firing consecutively.

\begin{figure}[htp]
    \centering
    \includegraphics[width=7cm]{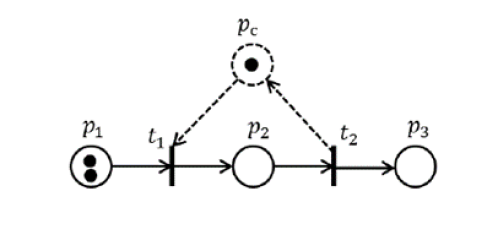}
    \caption{Control place in Petri nets}
    \label{fig:c.png}
\end{figure}

Ideally, the supervisor should only block those inputs that would inevitably lead to a violation of the desired specifications. When this occurs, the supervisor is considered \textit{maximally permissive}. This scenario is achievable when all transitions within the network are controllable; however, such circumstances are rare in practice.

It is assumed that the plant represents an initial network that requires a supervisory observer to ensure the closed-loop Petri net satisfies the desired specifications. The plant is modeled using \(n\) places and \(m\) transitions, with its incidence matrix denoted by \([N_p]\). Consequently, the incidence matrix of the closed-loop Petri net is expressed as:

\begin{equation}
[N] = 
\begin{bmatrix}
N_p \\ 
N_c 
\end{bmatrix}
\end{equation}

Here, \(N_c\) represents the supervisory control structure, comprising \(n_c\) distinct places. Importantly, the number of transitions in the network remains unchanged. In general, each GMEC can be mathematically expressed as follows:

\begin{equation}
(l, \beta) \equiv l^T M \leq b = \sum_{i=1}^{n} a_i^T m(p_i) \leq b
\end{equation}

where \(m(p_i)\) indicates the number of tokens at the \(i\)-th place in mode \(M\). Additionally, \(a_i\) (for all \(i \in N_n\)) are non-negative coefficients, \(b \in N\) is a fixed scalar, and \(l = (a_1, a_2, \dots, a_n)\).

Based on relation (17), network modes can be categorized into two types: acceptable modes (\(M_A\)) and forbidden modes (\(M_F\)), with forbidden modes violating the above inequality. Thus, the set of all system states is given by:

\begin{equation}
R(N, M_0) = M_A \cup M_F
\end{equation}

By considering all possible system states, inequality (17) can be expanded as follows:

\begin{equation}
\begin{cases} 
\sum_{i=1}^{n} a_i^T m(p_i) \leq b, & M \in M_A \\ 
\sum_{i=1}^{n} a_i^T m'(p_i) \geq b + 1, & M' \in M_F
\end{cases}
\end{equation}

This equation implies that the unknown coefficients \(a_i\) and \(b\) for each constraint must be chosen to eliminate the forbidden states from the reachability graph. Simply put, the objective of applying these constraints is to prevent the network from reaching states that do not satisfy (17). To achieve this, a monitor place corresponding to each forbidden state is added, transforming inequality (17) into the following equality:

\begin{equation}
\sum_{i=1}^{n} a_i^T m(p_i) + m(p_c) = b
\end{equation}

Relation (18) indicates that the monitor \(p_c\) forms a constant vector \(p\) with places where \(a_i \neq 0\). It is also evident that the initial marking of the monitor place will be negative.

Considering the \(n_c\) conditions necessary for complete system into account, the supervisory control structure can be determined using the following equation:

\begin{equation}
[N_c] = -L \cdot [N_p]
\end{equation}

The initial marking of the monitor places can also be calculated as follows:

\begin{equation}
M_{c_0} = B - L \cdot M_0
\end{equation}

Here, matrix \(L\) represents the non-negative coefficients for each constraint, while matrix \(B\) contains the fixed coefficients of the constraints:

\begin{equation}
L = 
\begin{bmatrix} 
l_1 & l_2 & \dots & l_{n_c} 
\end{bmatrix}, \quad 
B = 
\begin{bmatrix} 
b_1, b_2, \dots, b_{n_c} 
\end{bmatrix}^T
\end{equation}

In summary, GMECs can be classified into two groups: dependent constraints and independent constraints. By accurately identifying the independent constraints, the designed monitor places will also cover the dependent constraints, ensuring full control of the system.

\section*{Supervisory control algorithm using multiple model control systems}

When resources are unreliable, their failure can prevent dependent subsystems from functioning properly. In extreme cases, the failure of a single resource may result in the collapse of the entire system. To address these challenges\cite{b14}, researchers have developed several methods to enhance system robustness. Among these are recovery subnets, which model the processes of resource failure and recovery. Additionally, techniques such as minimal siphon detection and the development of GMECs are employed to effectively manage unreliable resources. These methods help ensure stable system performance by avoiding deadlocks and facilitating recovery from resource failures. Moreover, integrating these strategies significantly enhances network flexibility and reliability, thereby enabling more efficient and resilient manufacturing processes.

\subsection*{Assumptions}

Before outlining the steps for designing robust supervisory control, the key assumptions underlying this research must be reviewed:

\begin{itemize}
    \item \textbf{First assumption:} This study focuses on identifying forbidden states caused by resource failures. It is assumed that, in the initial state where all resources are functioning properly, the characteristic of liveness is maintained. Furthermore, resource failures do not compromise the system's boundedness.
    \item \textbf{Second assumption:} To avoid excessive computational complexity, the proposed algorithm is limited to addressing only one type of unreliable resource within the network.
    \item \textbf{Third assumption:} Unreliable resources can fail only when in an idle state. The idle or waiting state refers to a condition in which a resource is operational but not actively engaged in any task. Within the Petri net framework, the number of tokens at a specific resource location in each state represents the number of idle resource units.
\end{itemize}

\subsection*{Steps for Ensuring Liveness in Automated manufacturing Systems}

The steps required to ensure the liveness characteristic in automated manufacturing systems are as follows:

\begin{enumerate}
    \item Design the recovery subnet for unreliable resources.
    \item Analyze the impact of unreliable resources on network states.
    \item Apply a control mechanism to the network.
\end{enumerate}

When a resource failure occurs, the recovery subnet is triggered, guiding the system through a predefined process to resolve the issue. Assuming $N$ represents an S\textsuperscript{4}PR network, the set of resources $P_R$ is divided into two categories: reliable resources ($P_R^r$) and unreliable resources ($P_R^u$), where $P_R^r \cap P_R^u = \emptyset$. Consequently, the recovery subnet for an unreliable resource in a Petri net is defined as $(N_r, M_0^r) = \{(r, q), \{t_f, t_r\}, F, W, M_0^r\}$, where $r \in P_R^u$, $q$ represents the recovery state, $t_f$ signifies resource failure, and $t_r$ denotes resource recovery. The edges of this subnet are $F = \{(r, t_f), (t_f, q), (q, t_r), (t_r, r)\}$.

\subsubsection*{Step 1: Recovery Subnet Design}

Within the Petri net framework, when a failure occurs in an unreliable resource, the transition $t_f$ fires, moving the signal from the resource location to the recovery state. This indicates the resource has been removed for repair. Once the issue is resolved, the transition $t_r$ fires, returning the signal to the resource location. This process confirms that the repaired resource is ready to resume normal operation. It should be noted that the transitions in this subnet are uncontrollable and fire automatically.

\subsubsection*{Step 2: Impact on Network Behavior}

Resource failure can alter the overall behavior of the network by halting or partially executing certain processes, which may negatively affect system performance. A crucial method for addressing this is to analyze system stability. The objective is to assess the system's ability to withstand resource failures and continue functioning without significant disruption. This analysis categorizes system states into robust states ($M_R$) and non-robust states ($M_{UR}$) such that $R(N, M_0) = M_R \cup M_{UR}$.

A robust state ensures that, even in the event of an unreliable resource failure, subnets relying only on reliable resources can operate without interruption. Within Petri nets, a state or marking is considered robust if there exists a finite, non-zero firing sequence (e.g., $\alpha$) that allows the system to transition from its current state to these subnets. Otherwise, the state is deemed non-robust.

One challenge in distinguishing robust and non-robust states is identifying all system states. To address this, \cite{b5} proposes three general approaches:
\begin{enumerate}
    \item Performing a complete computation of the reachability graph to identify robust states through a firing sequence.
    \item Computing a subset of the reachability graph and repeating the first method.
    \item Developing a linear mathematical algorithm.
\end{enumerate}

Given that the number of network states increases exponentially with network size, calculating the reachability graph is often impractical. For small networks, the first two methods are preferable, whereas the third method is suitable for larger and more complex systems. This research focuses solely on the modes identified using one of these methods, as robust states cannot be included in the set of forbidden modes. It is also important to note that if a failure occurs in a non-robust state, the network will reach a deadlock. Therefore, the controller is designed to act only when a failure arises in a robust state.

\subsubsection*{Step 3: Robust Control Design}

The primary objective in designing robust control is to create a system capable of tolerating model errors and imperfections while still delivering the desired results \cite{b8}. For systems with shared resources such as S\textsuperscript{4}PR, a forbidden-mode controller is considered robust to resource failures if and only if two conditions are met:
\begin{enumerate}
    \item Under normal conditions, without resource failures, the network operates stably and continuously.
    \item In the event of an unreliable resource failure, the remaining resources are distributed among the available subnets to prevent the system from reaching a deadlock.
\end{enumerate}

To achieve this, the proposed algorithm is presented in this research.
\RestyleAlgo{ruled}
\begin{algorithm}
\caption{Optimal Controlled Petri Net Algorithm}\label{alg:optimal}
\KwData{S\textsuperscript{4} PR Petri Net $(N, M_0)$ with Robust and Unrobust markings $(M_R$ and $M_{UR})$ divided like Algorithm used in \cite{b5}.}
\KwResult{An optimal controlled Petri net corresponding to the capacity of an unreliable resource $(r)$.}
\For{$r \in P_R^u$}{
    Design recovery subnet: $(N_r, M_{0r}) = (\{r, q\}, \{t_f, t_r\}, F, w, M_{0r})$\;
    Set $M_{0r} = n$ and $M_{0q} = 0$\;
    \eIf{$M_q = n$}{
        \For{$N_i$ such that $P_{Ri} \cap P_R^u \neq \emptyset$, $i \in N_n$}{
            Construct $Model_n$ by adding inhibitor arc from $q$ to $\{t \in T \mid t \cap P_i^0{}^{\bullet} \neq \emptyset, i \in N_n\}$\;
        }
    }{
        Consider $M_q = j$\;
        \For{$j = 1:1:n-1$}{
            Calculate $[N_C]_j$\;
            Construct $Model_j$ by adding $[N_C]_j$ to $(N, M_0)$\;
        }
    }
}
\end{algorithm}

According to \cite{b5}, a firing sequence exists such that:
\begin{equation}
M_A = M_R
\end{equation}
Consequently, among unrobust markings, only those satisfying the following relation are considered forbidden markings:
\begin{equation}
\forall r \in P_R^u, \, M \in M_{UR} \rightarrow M_F = [M \mid m(r) \geq i]
\end{equation}
Which results in:
\begin{equation}
M_F \subseteq M_{UR}
\end{equation}
Forbidden markings are equal to or a subset of unrobust markings. This is because, in robust markings, there is at least one infinite firing sequence, denoted as \(\alpha\), that drives the network to independent sub-nets.

\textbf{Definition:}  
A marking \(m'\) is called a root marking of \(m\) (or \(m\) is a mask marking of \(m'\)) if and only if:
\begin{equation}
\forall p \in P_E \rightarrow m'(p) \geq m(p)
\end{equation}
It has been demonstrated in \cite{b9} that if the state \(m'\) is admissible, then \(m\) is also admissible. Moreover, if \(m\) is a forbidden marking, \(m'\) will also be a forbidden marking. Thus, root and mask markings can be excluded from the sets of forbidden and admissible markings. As a result, reduced sets of forbidden markings (\(M_{Fnew}\)) and admissible markings (\(M_{Anew}\)) are obtained.

The extended equation becomes:
\begin{equation}
\begin{cases}
\sum_{i \in N_A} a_i \cdot m''(p_i) \leq b, \, m'' \in M_{Anew} \\
\sum_{i \in N_A} a_i \cdot m'(p_i) \geq b + 1 - Q(1 - f_s), \, m' \in M_{Fnew}
\end{cases}
\end{equation}
Here, \(f_s \in \{0, 1\}\) is a binary variable defined for each constraint. \(f_s = 1\) indicates that the condition \(s\) is satisfied by the resulting monitor place. \(Q\) represents a very large positive number. If \(f_s = 0\), the constraint \(s\) is removed from the equations.

Solving these inequalities yields multiple solution sets, each defining the number of required monitor places. Ultimately, by applying the relationship provided in previous section, the appropriate supervisory control structure is derived.

Our goal is to construct a structure in which various resource failures transition the system to different states.

The recovery sub-net includes a control switch, comprising an action place and a switch place.
\begin{itemize}
    \item The \textbf{action place} selects the optimal algorithm based on the tokens in \(Q\).
    \item The \textbf{switch place} replaces the algorithm with the initial model.\cite{12}
\end{itemize}

\section*{Experimental Results}

In this section, we run our algorithm on an S\textsuperscript{4}PR net. To better understand the content, we consider an S\textsuperscript{4}PR network consisting of three sub-nets. Each sub-network represents a distinct production line that uses the resources available in the network.

\begin{figure}[htp]
    \centering
    \includegraphics[width=9cm]{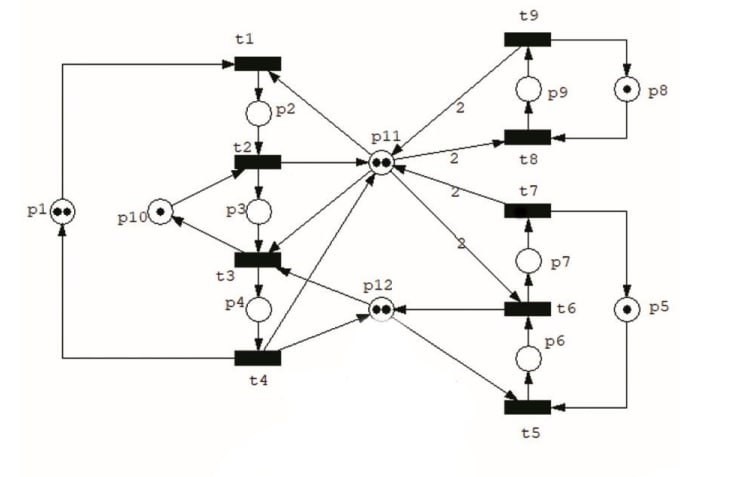}
    \caption{The S\textsuperscript{4}PR network we study our algorithm on}
    \label{fig:c.png}
\end{figure}

\begin{equation}
\begin{cases}
P_0 = \{p_1, p_5, p_8\}, \quad \\P_R = \{p_{10}, p_{11}, p_{12}\}, \quad \\P_A = \{p_2, p_3, p_4, p_6, p_7, p_8\}
\end {cases}
\end{equation}

The first subnet has the ability to process two parts simultaneously:
\begin{equation}
M_0^1 = 2 \cdot p_1
\end{equation}

The second subnet has the ability to process one piece at a time:
\begin{equation}
M_0^2 = p_5
\end{equation}

Finally, the third subnet has the ability to process one piece:
\begin{equation}
M_0^3 = p_8
\end{equation}

In this network, we consider place \( p_{12} \) as an unreliable resource. As seen, the capacity of this resource in the network is equal to two (\( c = 2 \)), and the processing operation in the third sub-network is independent of this resource. As a result:
\begin{equation}
P_R^u = \{p_{12}\}, \quad P_R^r = \{p_{10}, p_{11}\}
\end{equation}
The initial marking is:
\begin{equation}
M_0 = 2 \cdot p_1 + p_5 + p_8 + p_{10} + 2 \cdot p_{11} + 2 \cdot p_{12}
\end{equation}

The reachability graph of the network in its normal state is shown, containing 23 reachable markings. The liveness and boundedness of the system are guaranteed.
\begin{figure}[htp]
    \centering
    \includegraphics[width=9cm]{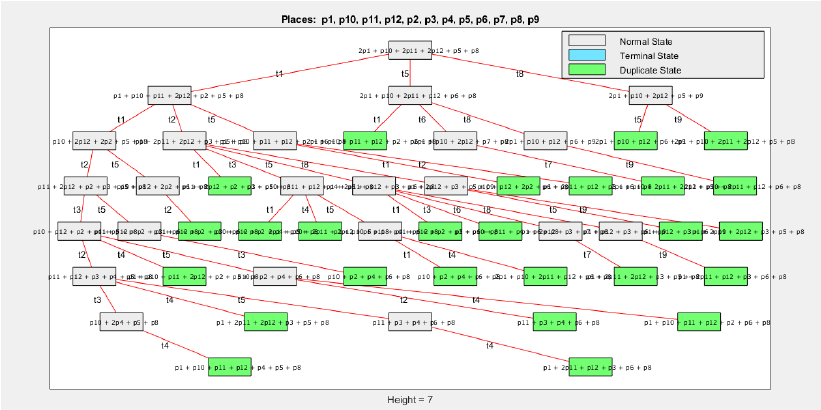}
    \caption{Reachablity graph of studied Petri net}
    \label{fig:c.png}
\end{figure}

Next, we update the network model by adding the recovery subnet defined in (Fig. 4.), which aims to monitor the status of the unreliable resource. As seen, place \( q \) is initially unmarked, indicating that the health of both capacities of the unreliable resource is unknown. As soon as a failure occurs, transition \( t_f \) will automatically fire, transferring the signal from the unreliable source place to the recovery place. Based on how many signs are placed in \( q \), the controller will replace the corresponding model to maximize network efficiency. Finally, after repairing the failure, transition \( t_r \) is automatically activated, transferring the signal from place \( q \) to place \( p_{12} \), and the original model will dominate the network again.
\begin{figure}[htp]
    \centering
    \includegraphics[width=5cm]{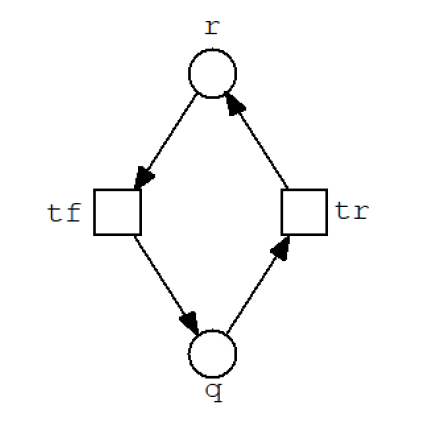}
    \caption{Recovery subnet}
    \label{fig:c.png}
\end{figure}

Before designing the controller, we examine the effect of the number of failed capacities on the reachability graph.

\subsection*{First Scenario: Both Signs in \( p_{12} \) Transferred to Recovery}
In this case, both signs in the source \( p_{12} \) are transferred to the recovery location for repair (\( i = 2 = n \)). The number of reachable markings will reduce to 7. This is predictable according to the original model since the absence of a token in \( p_{12} \) causes the second sub-network to never be activated, the first sub-network to run incompletely, and only the third sub-network (independent of \( p_{12} \)) to run completely and infinitely. Thus, the control objective in this scenario is to remove the subnets dependent on the unreliable resource so that only the third subnet continues to function.

\subsection*{Second Scenario: Only One Capacity from an Unreliable Source Withdrawn for Repair}
In this case, only one capacity from an unreliable source is removed from the network and placed in the recovery site for repair (\( i = 1 < n \)). The number of reachable markings decreases from 23 to 19 when a capacity is withdrawn from an unreliable resource. Additionally, although the production process does not stop in any subnet, the processing speed on the parts will be slower. The reduction in resource capacity has also led to a deadlock in the network.

The first step in this direction will be the correct identification of forbidden states. According to the explanations in (36) and the implementation of one of the three mentioned methods, the main network modes are divided into robust and unrobust categories:
\begin{equation}
R(N, M_0) = M_R \cup M_{UR}
\end{equation}
Where:

\begin{equation}
M_R = 
\footnotesize
\begin{bmatrix}
2 & 0 & 0 & 0 & 1 & 0 & 0 & 1 & 0 & 1 & 2 & 2 \\
2 & 0 & 0 & 0 & 1 & 0 & 0 & 0 & 1 & 1 & 0 & 2 \\
2 & 0 & 0 & 0 & 0 & 1 & 0 & 1 & 0 & 1 & 2 & 1 \\
1 & 1 & 0 & 0 & 1 & 0 & 0 & 1 & 0 & 1 & 1 & 2 \\
2 & 0 & 0 & 0 & 0 & 1 & 0 & 0 & 1 & 1 & 0 & 1 \\
2 & 0 & 0 & 0 & 0 & 0 & 1 & 1 & 0 & 1 & 0 & 2 \\
1 & 1 & 0 & 0 & 0 & 1 & 0 & 1 & 0 & 1 & 1 & 1 \\
1 & 0 & 1 & 0 & 1 & 0 & 0 & 1 & 0 & 0 & 2 & 2 \\
1 & 0 & 1 & 0 & 0 & 1 & 0 & 1 & 0 & 0 & 2 & 1 \\
1 & 0 & 1 & 0 & 1 & 0 & 0 & 0 & 1 & 0 & 0 & 2 \\
1 & 0 & 1 & 0 & 0 & 1 & 0 & 0 & 1 & 0 & 0 & 1 \\
1 & 0 & 1 & 0 & 0 & 0 & 1 & 1 & 0 & 0 & 0 & 2
\end{bmatrix}
\end{equation}

\begin{equation}
M_{UR} = 
\footnotesize
\begin{bmatrix}

0 & 2 & 0 & 0 & 1 & 0 & 0 & 1 & 0 & 1 & 0 & 2 \\
0 & 2 & 0 & 0 & 0 & 1 & 0 & 1 & 0 & 1 & 0 & 1 \\
1 & 0 & 0 & 1 & 1 & 0 & 0 & 1 & 0 & 1 & 1 & 1 \\
0 & 1 & 1 & 0 & 1 & 0 & 0 & 1 & 0 & 0 & 1 & 2 \\
1 & 0 & 0 & 1 & 0 & 1 & 0 & 1 & 0 & 1 & 1 & 0 \\
0 & 1 & 1 & 0 & 0 & 1 & 0 & 1 & 0 & 0 & 1 & 1 \\
0 & 1 & 0 & 1 & 1 & 0 & 0 & 1 & 0 & 1 & 0 & 1 \\
0 & 1 & 0 & 1 & 0 & 1 & 0 & 1 & 0 & 1 & 0 & 0 \\
0 & 0 & 1 & 1 & 1 & 0 & 0 & 1 & 0 & 0 & 1 & 1 \\
0 & 0 & 1 & 1 & 0 & 1 & 0 & 1 & 0 & 0 & 1 & 0 \\
0 & 0 & 0 & 2 & 1 & 0 & 0 & 1 & 0 & 1 & 0 & 0
\end{bmatrix}
\end{equation}

The recovery subnet is designed, and the following relations govern the system:
\begin{equation}
(N_r, M_{0_r}) = \{(p_{12}, q), (t_f, t_r), F, M_{0_r}\}
\end{equation}
\begin{equation}
F = \{(p_{12}, t_f), (t_f, q), (q, t_r), (t_r, p_{12})\}
\end{equation}

\subsection*{Second Step: Normal Mode}
In this case, \( i = 2 \), and in normal mode, \( i = 0 \).

\subsection*{Third Step: \( i = n \)}
Two restraining arcs will be recovered from the place and connected to the first and second subnets with a weight of two.

\subsection*{Fourth Step: \( i = n - 1 = 1 \)}
\begin{equation}
\footnotesize
M_A = M_R = \begin{bmatrix}
2 & 0 & 0 & 0 & 1 & 0 & 0 & 1 & 0 & 1 & 2 & 2 \\
2 & 0 & 0 & 0 & 1 & 0 & 0 & 0 & 1 & 1 & 0 & 2 \\
2 & 0 & 0 & 0 & 0 & 1 & 0 & 1 & 0 & 1 & 2 & 1 \\
1 & 1 & 0 & 0 & 1 & 0 & 0 & 1 & 0 & 1 & 1 & 2 \\
2 & 0 & 0 & 0 & 0 & 1 & 0 & 0 & 1 & 1 & 0 & 1 \\
2 & 0 & 0 & 0 & 0 & 0 & 1 & 1 & 0 & 1 & 0 & 2 \\
1 & 1 & 0 & 0 & 0 & 1 & 0 & 1 & 0 & 1 & 1 & 1 \\
1 & 0 & 1 & 0 & 1 & 0 & 0 & 1 & 0 & 0 & 2 & 2\\
1 & 0 & 1 & 0 & 0 & 1 & 0 & 1 & 0 & 0 & 2 & 1 \\
1 & 0 & 1 & 0 & 1 & 0 & 0 & 0 & 1 & 0 & 0 & 2\\
1 & 0 & 1 & 0 & 0 & 1 & 0 & 0 & 1 & 0 & 0 & 1 \\
1 & 0 & 1 & 0 & 0 & 0 & 1 & 1 & 0 & 0 & 0 & 2
\end{bmatrix}
\end{equation}
As we have:
\begin{equation}
M_F \subseteq M_{UR} 
\end{equation}
So:
\begin{equation}
\footnotesize
 M_F = \begin{bmatrix}
0 & 2 & 0 & 0 & 1 & 0 & 0 & 1 & 0 & 1 & 0 & 2 \\
0 & 2 & 0 & 0 & 0 & 1 & 0 & 1 & 0 & 1 & 0 & 1 \\
1 & 0 & 0 & 1 & 1 & 0 & 0 & 1 & 0 & 1 & 1 & 1 \\
0 & 1 & 1 & 0 & 1 & 0 & 0 & 1 & 0 & 0 & 1 & 2 \\
1 & 0 & 0 & 1 & 0 & 1 & 0 & 1 & 0 & 1 & 1 & 0 \\
0 & 1 & 1 & 0 & 0 & 1 & 0 & 1 & 0 & 0 & 1 & 1 \\
0 & 1 & 0 & 1 & 1 & 0 & 0 & 1 & 0 & 1 & 0 & 1 \\
0 & 1 & 0 & 1 & 0 & 1 & 0 & 1 & 0 & 1 & 0 & 0 \\
0 & 0 & 1 & 1 & 1 & 0 & 0 & 1 & 0 & 0 & 1 & 1 \\
0 & 0 & 1 & 1 & 0 & 1 & 0 & 1 & 0 & 0 & 1 & 0 \\
0 & 0 & 0 & 2 & 1 & 0 & 0 & 1 & 0 & 1 & 0 & 0
\end{bmatrix}
\end{equation}

Now, we will implement the size reduction algorithm of the set of admissible and forbidden markings:
\begin{equation}
M_{Ar} = \begin{bmatrix}
0 & 0 & 0 & 0 \\
0 & 0 & 0 & 0 \\
0 & 0 & 0 & 1 \\
1 & 0 & 0 & 0 \\
0 & 0 & 0 & 1 \\
0 & 0 & 0 & 0 \\
1 & 0 & 0 & 1 \\
0 & 1 & 0 & 0 \\
0 & 1 & 0 & 1 \\
0 & 1 & 0 & 0 \\
0 & 1 & 0 & 1 \\
0 & 1 & 0 & 0
\end{bmatrix}
\end{equation}

\begin{equation}
M_{Fr} = \begin{bmatrix}
2 & 0 & 0 & 0 \\
2 & 0 & 0 & 1 \\
0 & 1 & 0 & 0 \\
0 & 2 & 0 & 0 \\
0 & 1 & 0 & 1 \\
0 & 0 & 1 & 1 \\
0 & 0 & 0 & 0
\end{bmatrix}
\end{equation}

\begin{figure}[htp]
    \centering
    \includegraphics[width=10cm]{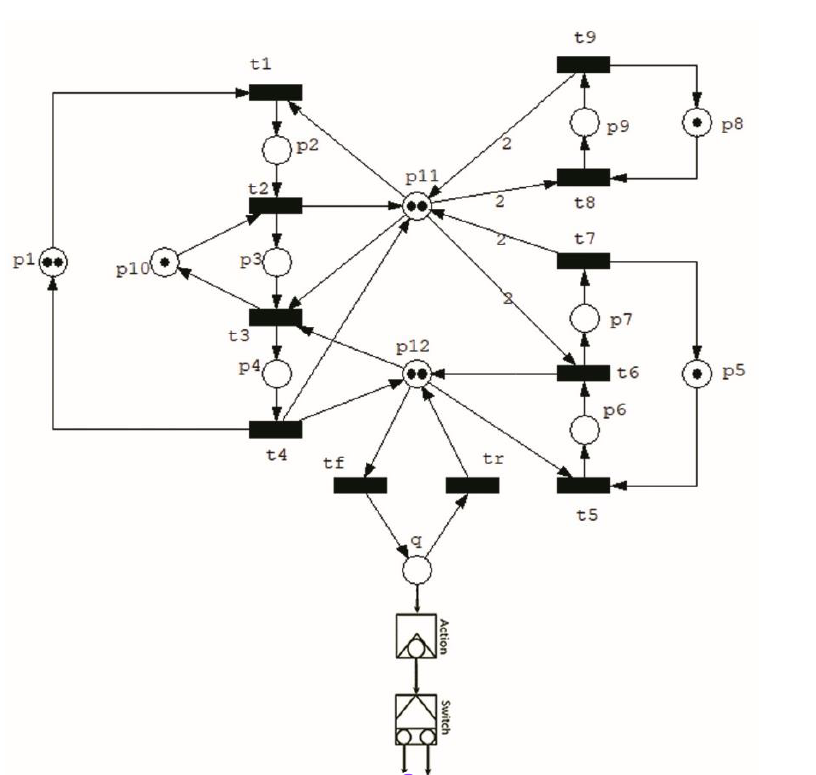}
    \caption{final controlled model}
    \label{fig:c.png}
\end{figure}

\section*{Conclusion}

Petri nets are powerful graphical and mathematical tools for modeling dynamic systems, particularly discrete event systems. When monitoring system’s performance, liveness is essential since a lack of liveness can lead to deadlock. From a behavioral perspective, deadlocks typically arise from two main issues: improper allocation of available resources or physical damage to the resources. The impact of resource failures on network states has received limited attention. The primary objective of this paper is to enhance system productivity and production rates by reallocating the network's remaining resources. Adding control places using the GMEC method offers a practical solution by leveraging both the structural and behavioral properties of the network. This approach considers the remaining quantity of an unreliable resource in the network while identifying admissible and forbidden states at each stage. The outcome of this process is the development of a new controller algorithm. Finally, a multi-model structure within the Petri net framework is proposed, where a pre-designed model is replaced with one that accounts for the remaining quantity of a reliable resource, substituting any non-functional model. By modeling systems to ensure the inclusion of features essential for optimizing plant operations, this approach mitigates operational disruptions and minimizes significant financial losses associated with equipment damage and wasted raw materials on the production line.

\end{document}